\documentstyle[12pt]{article}
\begin{document}    
\thispagestyle{empty}

\begin{center}\LARGE{The X-ray Structure of 
A399 and A401: A Pre-Merging Cluster Pair 

}
\end{center}

\bigskip

\begin{center}\Large
{Yutaka Fujita$^{1,2}$, Katsuji Koyama$^{3}$, Takeshi Tsuru$^{3}$, 
and Hironori Matsumoto$^{3}$ }
\end{center}

\begin{itemize}\large\sl
\item[] ${}^{1}$Graduate School of Human and Environmental Studies\\
Kyoto University, Kyoto 606-01, Japan
\item[] ${}^{2}$Yukawa Institute for Theoretical Physics\\
Kyoto University, Kyoto 606-01, Japan
\end{itemize}

\begin{center}\sl
E-mail(YF) fujita@yukawa.kyoto-u.ac.jp
\end{center}

\begin{itemize}\large\sl
\item[] ${}^{3}$Department of Physics, Faculty of Science\\ 
Kyoto University, Kyoto 606-01, Japan\\ 
\end{itemize}

\bigskip
\noindent
{\em Subject headings}: Galaxies: Clustering - Galaxies: clusters:
individual: (A399, A401)
- X-rays: Galaxies

\abstract

{We present {\em ASCA} results  of the pair clusters A399 and 
A401. The region between the two clusters exhibits excess    
X-rays over the value expected with a simple superposition of the two
clusters. 
We see, however, 
no temperature rise at the merging front; the temperature is near 
the average of those  in the inner regions of the two clusters.
These indicate that the two clusters are really interacting but
it is not strong at present.
 The inner regions of the two clusters show no radial 
variations of temperature, 
abundance and absorption values.
We set upper-limits of mass deposition rate of cooling flow to be  
$\dot{M}<35\;\rm M_{\odot}yr^{-1}$ and 
$\dot{M}<59\;\rm M_{\odot}yr^{-1}$ for A399 and A401, respectively. 
A  hint of azimuthal variation
of the temperature is also found.}

\newpage

\section{INTRODUCTION}
\label{sec-1}
\baselineskip=20pt
\indent

  A  large fraction of clusters of galaxies have been found to contain
substructures in the optical galaxy distributions and X-ray surface 
brightness. This may be understood in the framework of the 
hierarchical clustering 
scenario; clusters of galaxies evolve by accreting other clusters or 
smaller groups of 
galaxies.
  By the merging, the volume or density of the X-ray emitting 
gas may increase, hence X-ray luminosity  
increases as clusters evolve. Gioia et al.(1990) and 
Edge et al. (1990), for example, reported  evidence for the 
luminosity evolution of the  X-ray emitting intracluster gas.
 Recent observations with {\em ROSAT} provide more direct 
evidence for the cluster merging; in some clusters, 
their X-ray 
surface brightness distributions are shifted from the galaxy 
number density distribution, 
in which local high temperature regions are found. These 
indicate that the clusters are actually merging 
with a strong shock at the merging front (Henry \& Briel 1995 ; 
Zabludoff \& Zaritsky 1995 ; Burns et
al. 1995). 
 Suppose that the subcluster components are more separated, 
then the cluster may not be regarded as a single 
cluster
but would be counted as two separate clusters. The 
distinction may be somewhat ambiguous; it depends on the
projected distance of the clusters.  If the system is gravitationally 
bounded, then the clusters will merge and 
evolve to a single, richer cluster in the future. 

The pair of A399 and A401 
is one of the closest and the
brightest pair clusters, hence would be the best candidate of pre-merging
clusters. Karachentsev and Kopylov (1980) measured the velocity
dispersions and mean radial velocities of the two clusters and
concluded that these data are consistent with the pair being
gravitationally bound. However, although several studies have 
been made on the pair 
clusters, whether they are thermally interacting or not is still
open problem.

Some authors claimed that they may be interacting.
McGlynn \& Fabian (1984) pointed out that their lack of cooling flows 
may be due to the interaction.
Although both of these two clusters possess luminous cD galaxies, 
their cooling flows are 
relatively weak (McGlynn \& Fabian 1984). By numerical
simulations of cluster collisions, they confirmed 
that, at the first encounter, 
one cluster can even pass through the other without destroying the
optical components, while the intracluster gas is largely 
affected and hence cooling flows, if any, may be disrupted
(McGlynn \& Fabian 1984). Furthermore, A401 is one of only a few rich
clusters that is report to have an extensive cluster radio halo
(Harris, Kapahi, \& Ekers. 1980 ; Roland et al. 1981). Harris et al. (1980)
suggested that radio halos form during the coalescence of clusters.

On the other hand, X-ray observation with the {\em Einstein} satellite 
indicated that the emission between the two clusters is explained by the 
simple superposition of the extended emission from the two clusters (Ulmer \&
Cruddace 1981).  

We report, for the first time, spatially resolved X-ray spectra of the
pair A399 and A401 with the {\em ASCA} satellite.
We found evidence for the interaction between these two clusters; 
there is an excess X-ray emission in the middle of the two 
clusters above the   
expected flux from the simple superposition of the two clusters.
We also discuss  large-scale temperature variations in the outer regions of
the two clusters, which may be affected by the interaction between 
them.

We describe the observations and the data reduction in section 2 and the
analysis and the results in section 3. Then, in section 4, we discuss the
structure of the two clusters and the interaction between them 
based on the results. Section 5 is
devoted to conclusions. 
We assume $H_{0}=50 {\rm km s^{-1} Mpc^{-1}}$ and $q_{0}=0.5$ through 
out this paper.

\section{Observations and Data reduction }
\label{sec-2}
\indent

 We observed A399, A401 and the region between the two clusters 
(hereafter we call it the link-region) 
with the {\em ASCA} satellite on 16, 18 and 17 August 1994, respectively. 
We rejected data at low elevation angles from bright earth
($<20^{\circ}$) and those affected by South Atlantic Anomaly. Hot and
flickering pixels for the SIS data were removed. The rise-time      
rejection technique was used to reject particle events in the GIS   
data. 
 After these data screening, the accumulation times for A399,         
A401 and the
link-region are $2.6 \times 10^{4}$s, $2.9 \times 10^{4}$s and $3.5
\times 10^{4}$s, respectively. 
 Figure 1 shows GIS image of the two clusters; a mosaic map of the 
three pointing observations in which corrections for both the exposure 
time and telescope vignetting are made. 
 We divided the 
observed regions symmetrically
to the line connecting the centers of the two clusters as are  
illustrated in figure 1, 
and made separate spectra from each region. 
 Central regions of A401 and A399 (within $6'$ radius from the 
X-ray peak of each cluster) 
are denoted as A0 and B0, respectively, 
while the regions of C0 and D are located between the two clusters 
with radius of $6'$ and $10'$, respectively.
  The center positions of regions A0, B0, C0 and D are listed in table 1, 
together with
the optical centers (the positions of the cD galaxies: 
Hill \& Oegerle 1993) of A399 and A401. 
 Quadrants of the annulus around the regions of A0, B0 and C0, with 
$6'$ inner and $18'$
of outer radius are assigned as A1-A4, B1-B4 and C1-C2, respectively. 
We further  divided the inner  regions A0 and
B0 into 6 annulus with  $1'$ interval, and made separate spectra.
 The SIS spectra are only available for the regions of A0, B0 and D, 
because a large fraction 
of the other regions are outside of the SIS field of view, and because 
the poor sensitivity 
at the high energy of SIS makes it hard  to obtain 
reliable  wide-band X-ray spectra 
in the outer regions of the clusters.
The spectra in the series A and B were made using the data of the   
pointings centered on A399 and A401, respectively, while those in the series C and
D were derived from the
pointing data at the link-region.
The background spectra were made using the averaged blank field data provided 
by {\em ASCA} group,
from the same detector positions as the on-source spectra.
Then the  background subtractions were made after correcting the exposure time.
 For the image analysis, we have used an improved response function of {\em ASCA} telescope
provided by the XRT hardware team (Shibata, Honda, \& Hirayama, private communication), in 
which energy dependent point spread function and
effective area are taken into account.  In fact, the ray-tracing simulation demonstrated
that the error of the observed temperature is $< 10$\%  in the relevant temperature of 
5-10 keV (Shibata, Honda, \& Hirayama, private communication). Stray light from the two 
clusters
in region C0 is also estimated to be  $< 30$\% of the observed flux
in the region (Honda, private communication).

\section{Analysis and Results}
\label{sec-3}
\indent

The background subtracted spectra were binned to contain $> 20$ counts 
for each energy 
bin; minimum counts for which $\chi^{2}$ statistics can be applied, and then 
fitted to a thin thermal Raymond \& Smith model (1977) 
plus an interstellar absorption using the XSPEC package.
 The spectra from GIS2 and GIS3 or SIS0 and SIS1 were simultaneously fitted
with separate response functions and independent normalization factors 
for each detector.
  We show some examples of GIS and SIS results separately  
in figures 2 and 3; the GIS and SIS spectra and their best-fit model curves
in the regions of A0, B0 and C0 (for SIS, the spectrum is from the region D).
In all the regions, these models were found to be acceptable with the best-fit 
reduced $\chi^{2}$ values in the range of 0.8 - 1.1 for the GIS data and
1.0 - 1.5 for the SIS data.  The best-fit parameters are listed in 
tables 2 and 3. 

  The best-fit parameters obtained with  the GIS and  SIS spectra
agree 
well with each other, except for
absorption column densities, $N_{\rm H}$. These are 
derived essentially from the
spectral below  1 keV. In this energy band, SIS is more
sensitive than GIS, hence
the SIS results  would be more reliable.  
However, we notice here that 
calibration errors in the low energy band
give some bias to the estimated values. Therefore the 
absolute absorption may have systematic error, if it is less 
than a few times $10^{21} \rm cm^{-2}$, which is 
the case of the present observations. Nevertheless,  
we can reliably compare relative values of $N_{\rm H}$  
from the SIS. The relative values of $N_{\rm H}$ are almost 
constant from position to position. 
Therefore, we regard that the absolute  $N_{\rm H}$ is also 
constant everywhere in our fields.
 The $N_{\rm H}$ values determined with SIS are a few times 
larger than that of the galactic
absorption, which is $1.1\times 10^{21}\:\rm cm^{-2}$ (The Einstein 
On Line Service, Smithsonian Astrophysical Observatory). 
However, as we already noted, they are within the calibration errors, 
hence we do not discuss further for the $N_{\rm H}$ structure.

\subsection{Radial Structure}
\indent

 In figures 4 and 5, we show radial profiles of the
temperature of A399 and A401, respectively, in which inner $6'$ regions are 
divided to 6 annulus of $1'$ interval, while
the data of $6'-18'$ are averages of A1-A4  and of  B1-B4, for the 
A399 and A401, respectively.   
Uncertainty ranges of the fit 
given in the figures are 90\% confidence 
( hereafter errors are 90\% confidence
unless otherwise mentioned). 
 We see no significant radial variations of the temperature 
in  the inner regions.  
As already noted, we see no variations of the  $N_{\rm H}$ value 
(SIS data), neither. These indicate that  no
significant cooling flow components exist around the cD galaxies. 
  We further studied the SIS data, because SIS has 
better sensitivity than  GIS at the energy of relevant 
temperature of the cooling flows.
 In order to find upper limits of the cooling flows 
we fit the SIS spectra from the whole regions of A0 and B0 to a 
thin thermal plasma model (Raymond \& Smith 1977) 
plus an absorption. We have fixed the temperature, abundance
and absorption column density to their best-fit values and added a  
cooling flow component model as described in Mushotzky \& Szymkowiak 
(1988), in which the 
cooling is assumed to start from the temperature of the ambient intracluster 
gas. 
We thus estimate that the flow rates of the cooling gas are 
$\dot{M}<35\;\rm M_{\odot}yr^{-1}$ for A399 and
$\dot{M}<59\;\rm M_{\odot}yr^{-1}$ for A401. 
 
The average temperature in the central regions within $6'$ are 
found to be  6.9 keV and   8.0  keV, for A399 and A401, respectively
(figure 1, table 3). 
 Moving to the larger radius of $6'-18'$, the average temperatures
show slight increases for both the clusters;
from 6.9 keV to 8.1 keV and from 8.0 keV to 9.3 
keV, for A399 and A401, respectively. 
However the 90\% errors overlap with each other and 
the calibration of {\em     
ASCA} would be still incomplete. Thus we cannot argue the temperature
rise in the radial 
direction is real.  

 We investigated the abundance distributions of A399 and A401 as a 
function of radius, and 
found no significant  radial variations, although the errors are large. 
 
Thus, together with the global isothermality as discussed above,
the intracluster plasma near the cluster centers would be rather uniform.

\subsection{The Link-region} 
\indent

In order to see whether there are excess X-rays or not in the link-region, 
in figure 6, we plotted a photon flux  profile along the strip AA' 
with $5'$ width 
(see figure 1). 
  Assuming right-left symmetry of the flux distribution at the 
positions of cD galaxies
for each cluster, we show, with the dotted lines in figure 6,  
possible contributions of A401 and A399 to the link region.
 We clearly found excess X-rays in the 
link-region. 
 Further support of the excess emission is found
in the energy fluxes of regions  
A1, A3, B1 and B3 as are listed in 
table 3. The fluxes are of the best-fit parameters in 
table 3 for various regions. They are corrected for vignetting
and background. 
The sum of the energy fluxes of A3 and B3 is 
$F_{33} = 11.0 \times 10^{-12}\rm erg\: cm^{-2} s^{-1}$, 
and the sum of the energy fluxes of A1 and B1 is
$F_{11} = 6.3 \times 10^{-12}\rm erg\: cm^{-2} s^{-1}$. 
If the two clusters were spherically symmetric and they were simply
superposed on the line of sight, the sum of the expected energy flux 
contribution of A399 in B3 and that of A401 in A3 
would be $\tilde {F_{33}} \simeq 2^{-3} \times F_{11}$, 
because the X-ray surface brightness distribution in the outer
region of a typical cluster is well fitted by $r^{-3}$. 
 The fact $F_{33} > F_{11} + \tilde F_{33}$ indicates that the
the two clusters are physically interacting. 
 In figure 7, we present the best-fit temperature distribution
along the major axis.  The temperature shows slow decrease from  
A401 to A399, 
and that of the rink-region
is in the average value of A401 and A399.  
The abundance distribution is also examined, but is unclear due 
to its large errors.

\subsection{Non radial Structure} 
\indent

 The best-fit temperatures in the quadrants of the outer 
cluster regions and the average values in the inner regions 
( A0 and B0) are shown in figures 1.
We found that the temperature of the outer regions is, 
in average, higher than those in the inner regions in 
both the clusters, as already noted.  We further found a 
signature of non radial temperature variations in A399, although we
can reject the isothermal model at most at the 80\% confidence level.  
We note that the calibration errors in 
azimuthal direction are small compared with those in in radial
direction.  
As for the abundance, relative large errors prevent us to 
say definite statements. 

\section{DISCUSSION}
\label{sec-4}
\indent

The previous {\em Einstein} MPC results show that the 2 - 10 keV 
fluxes of A399 and A401 is $3.4 \times 10^{-11} \rm erg \: 
cm^{-2} s^{-1}$ and $5.9 \times 10^{-11} 
\rm erg \: cm^{-2} \: s^{-1}$, respectively, and their temperatures
are 
6.0 keV and 8.6 keV , respectively ( Edge et al. 1990 ; David et al. 1993). 
We determine the 2 - 10 keV fluxes of A399 and A401 within $18'$
(about 2 Mpc) to be  
$3.0 \times 10^{-11} \rm erg \: cm^{-2} s^{-1}$ and $5.7 \times 10^{-11} 
\rm erg \: cm^{-2} \: s^{-1}$, which is in excellent  
agreement to that determined with {\em Einstein}. 
  The mean temperatures determined with {\em ASCA} 
of about 6.9 and 8.0 keV agree also 
well to the {\em Einstein}.  
 The inferred temperature and luminosity fit well to the 
empirical correlation given by David et al.(1993).

 We found no significant radial variations in the central regions 
($< 6'$ = 0.8 Mpc) of  the clusters.
This, however,  does not immediately indicate that the radial 
structure of iron found within about 
100 kpc in the Centaurus and  Virgo  clusters is absent in A401 or A399, 
because the distance to the two clusters gives angular distance 
of $1'$, the spatial resolution of {\em ASCA}, to be 130kpc 
(Fukazawa et al. 1994 ; Koyama et al. 1991).
  We found no signature of cooling flows for both the clusters, 
giving upper limits of the 
mass deposition rates of $\dot{M}<35\;\rm M_{\odot}yr^{-1}$ for A399 and
$\dot{M}<59\;\rm M_{\odot}yr^{-1}$ for A401. 
These values are relatively small as for the clusters with cD galaxy 
(McGlynn \& Fabian 1984).

  The most important information of the present observation is 
the X-ray emission in the region between A401 and A399 clusters.  
 Ulmer and Cruddace (1981) claimed that the Einstein data can be explained as 
the simple superposition of foreground (A399) and background (A401) 
extended emission of the clusters.  We found clear excess X-rays in  
the link-region over the value 
expected with a simple superposition of the two clusters. 

 Numerical simulations of Evrard (1990) and Schindler \& M\"{u}ller (1993)
showed that during collision, gas between two clusters should be 
heated. This is not the case of our observation; the temperature in 
the link-region derived 
by {\em ASCA} is almost the same as those of A0 and B0. 
Furthermore,  table 1 shows that the locations of 
the X-ray peaks and the cD galaxies in each cluster 
coincide within the error ($\sim 1'$). 
 These facts support that interaction between A401 and A399 does exist, 
but no strong shock has not been made
in the link-region. Thus we interpret that the cluster pair A401-A399 
is currently in pre-merging phase.
 We guess that after several Gyr, the pair of the clusters 
will exhibit similar 
X-ray properties such as those of A754 and A2255, 
which are thought to be currently colliding
(Henry \& Briel 1995 ; Zabludoff \& Zaritsky 1995 ; Burns et al. 1995).
In A754, some regions have temperature of $T>12$keV while in
some other regions $T<6$keV.  In A754 and A2255, 
the locations of the brightest galaxies
in the clusters are away from the X-ray peaks more than $1'$ (Henry \&
Briel 1995; Burns et al. 1995).

 As for less possible scenario, although we cannot reject with 
the present result, is that 
the pair clusters is a relic of a past-strong collision.
A recent numerical simulation in which parameters are taken from the
observed values of A399 and A401  shows that after their 
collision, the temperature between 
the two clusters goes back to the initial value by an adiabatic expansion
in $\sim 1-2 \rm Gyr$ (Ishizaka 1995 private communication).
  The projection distance between A399 and A401 is 4.25Mpc 
(Karachentsev and Kopylov 1980) and their line-of-sight 
velocity separation, $c\Delta 
z$, is $\sim 1000 \rm km \: s^{-1}$. If we take these values 
as the real spatial
separation and relative velocity, the time needed for the two clusters
to separate as is observed 
is 4 Gyr and is well shorter than the Hubble time.  Even if $H_{0}=100
{\rm km \: s^{-1} \: Mpc^{-1}}$, the conclusion is the same.
Their lack of cooling flows may favor the past strong 
interaction; with the strong interaction
cooling flows were destroyed.

 The spectra in outer regions of A399 may show azimuthal variations,
although error bars are large. If they are true, 
these variations might be relics of the past-interaction between the
two clusters, or might be made by 
interactions of possible subclusters in the cluster.

The high temperature found in the region A4 may reflect 
the generation of shock waves caused by the collision between the
subclusters and the main clusters. However, figure 1 shows that there
are no significant substructures except for the extensions toward the
link-region. 

 An interpretation  can be made that the variances reflect the relics of the 
gravitational perturbations at the cluster formation period, which have not 
grown up as significant subclusters; dark matter blobs which correspond to the 
perturbations would have collided into the central regions of the clusters,
the gas trapped in them would have been shocked and  the temperature 
would have risen. 

\section{CONCLUSION}
\label{sec-5}
\indent

We report {\em ASCA} X-ray observation of the pair clusters A399 and
A401. No significant radial variations are found in the central
regions of the two clusters. There is an emission excess in the 
region between them 
(link-region) over the value expected with a simple superposition of the 
clusters. The temperature in it is not higher than
that in the central regions of the clusters. 
In the outer regions of A399, signatures of azimuthal
variation of temperature are found.

These results suggest that the two clusters are in a 
pre-merging phase; the interaction is not
strong at present.  We cannot exclude a  past strong collision.  
We encourage more
detailed observations and the comparison with numerical simulations.

\vspace{7mm}

We thank all of the members of the {\em ASCA} team and the launching staffs of
the Institute of Space and Astronautical Science. We thank C. Ishizaka 
for providing simulation data. 

\newpage

\section*{REFERENCES}

\newcommand\pp{\par\parshape 2 0truecm 13.5truecm 1truecm 12.5truecm\noindent}
\def\pap#1;#2;#3;#4; {\pp#1, {\sl #2}, {\bf #3}, #4.}
\def\book#1;#2;#3;#4; {\pp#1, {\sl #2} (#3: #4).}
\def\proc#1;#2;#3;#4;#5;#6;#7; {\pp#1, #2 {\sl #3}, #4; (#5: #6), #7.}
\def\ppt#1;#2; {\pp#1, #2.}

\pp Burns, J. O., Roettiger, K., Pinkney, J., Perley, R. A., Owen,
F. N., \& Voges, W. 1995,
ApJ.,  446, 583
\pp David, L. P., Slyz, A., Jones, C., Forman, W., \& Vrtilek, S. D. 1993,
ApJ.,  412, 479
\pp Edge, A. C., Stewart, G. C., Fabian, A. C., \& Arnaud, K. A. 1990,
Mon.Not R. astr.Soc.,  245, 559
\pp Evrard, A. E. 1990,
in Clusters of Galaxies, ed. W. R. Oegerle, M. J. Fitchett, \&
L. Danly (Cambridge Univ. Press), 287
\pp Fukazawa, Y., Ohashi, T., Fabian, A. C., Canizares, C.
R., Ikebe, Y., \mbox{Makishima,} K., Mushotzky, R. F., \& Yamashita, K. 1994,
Pub.Astr.Soc.Japan,  46, L55
\pp Gioia, I. M., Henry, J. P., Maccacaro, T., Morris, S. L., 
Stocke, J. T.,\& Wolter, A. 1990, 
ApJ., 356, L35
\pp Harris, D. E., Kapahi, V. K., \& Ekers, R. D. 1980,
Astron.Astrophys.Suppl., 39, 215
\pp Henry, J. P., \& Briel, U. G. 1995,
ApJ., 443, L9
\pp Hill, J. M., \& Oegerle, W. R. 1993,
AJ., 106, 831
\pp Karachentsev, I. D., \& Kopylov, A. I. 1980,
Mon.Not.R.astr.Soc., 192, 109
\pp Koyama, K., Takano, S., \& Tawara, Y. 1991,
Nature,  350, 135
\pp McGlynn, T. A., \& Fabian, A. C. 1984,
Mon.Not.R.astr.Soc., 208, 709
\pp Mushotzky, R. F., \& Szymkowiak, A. E. 1988,
in Cooling Flows in Clusters of Galaxies, ed. Fabian, A. C. (Kluwer
Academic Publishers), 53 
\pp Raymond, J. C., \& Smith, B. W. 1977,
ApJS. 35, 419
\pp Roland, J. H., Sol, H., Pauliny-Toth, I., \& Witzel, A. 1981,
Astron.Astrophys.,  100, 7
\pp Schindler, S., \& M\"{u}ller, E. 1993,
Astron.Astrophys.,  272, 137
\pp Ulmer, M. P., \& Cruddace, R. G. 1981,
ApJ., 246, L99
\pp Zabludoff, A. I., \& Zaritsky, D. 1995,
ApJ., 447, L21

\newpage
\thispagestyle{empty}

\section*{Figure Captions}

\noindent
 Fig.1. - The background-inclusive mosaic images taken with the GIS
 sensor, the configuration of the regions for spectral analysis (bold) 
and the best fit temperature (keV) observed with GIS (roman) and SIS (italic).
Exposure and vignetting area corrected. The contours are drawn at 5,
8, 13, 21, 33, 52 and 84 counts $\rm s^{-1}\:arcmin^{-2}$.

\noindent
 Fig.2. - X-ray spectra observed with two GIS. 
The histogram shows the best
fitting (R - S plasma + absorption column density).
(a) Region A0 (b) Region B0 (c) Region C0

\noindent
 Fig.3. - X-ray spectra observed with two SIS. 
The histogram shows the best
fitting (R - S plasma + absorption column density).
(a) Region A0 (b) Region B0 (c) Region D

\noindent
 Fig.4. - Profiles of temperature
  vs. the distance 
from the center of A399. 

\noindent
 Fig.5. - The same as Fig.5 but for A401.

\noindent
 Fig.6. - The solid lines show the photon counts after the correction of 
 exposure time and vignetting along the strip AA'    
(Fig.1). 
The dotted lines are reflection of the solid lines 
against to the center of the clusters. The dashed line indicates 
the background 
level. 

\noindent
 Fig.7. - Profiles of temperature along the line which connects A399
and A401.

\newpage
\thispagestyle{empty}

\begin{center}
TABLE 1 \\
LOCATIONS  (2000)
\end{center}

\begin{center}
\centering
\begin{tabular}{ c c c c }                \hline\hline
   &  z  & X-ray & $\rm cD^{a}$  \\ \hline
A0(A399)&0.0715&  2h57m53.25s, $+13^{\circ}02'39.26''$ & 
            2h57m53.15s, $+13^{\circ}01'51.9''$      \\
B0(A401)&0.0748&  2h58m58.97s, $+13^{\circ}34'57.35''$ &   
            2h58m57.88s, $+13^{\circ}34'59.1''$       \\
C0 and D &...&2h58m30.87s, $+13^{\circ}18'14.36''$ & 
            ...                                     \\  \hline

\end{tabular} 

\end{center}

a. Hills \& Oegerle 1993.

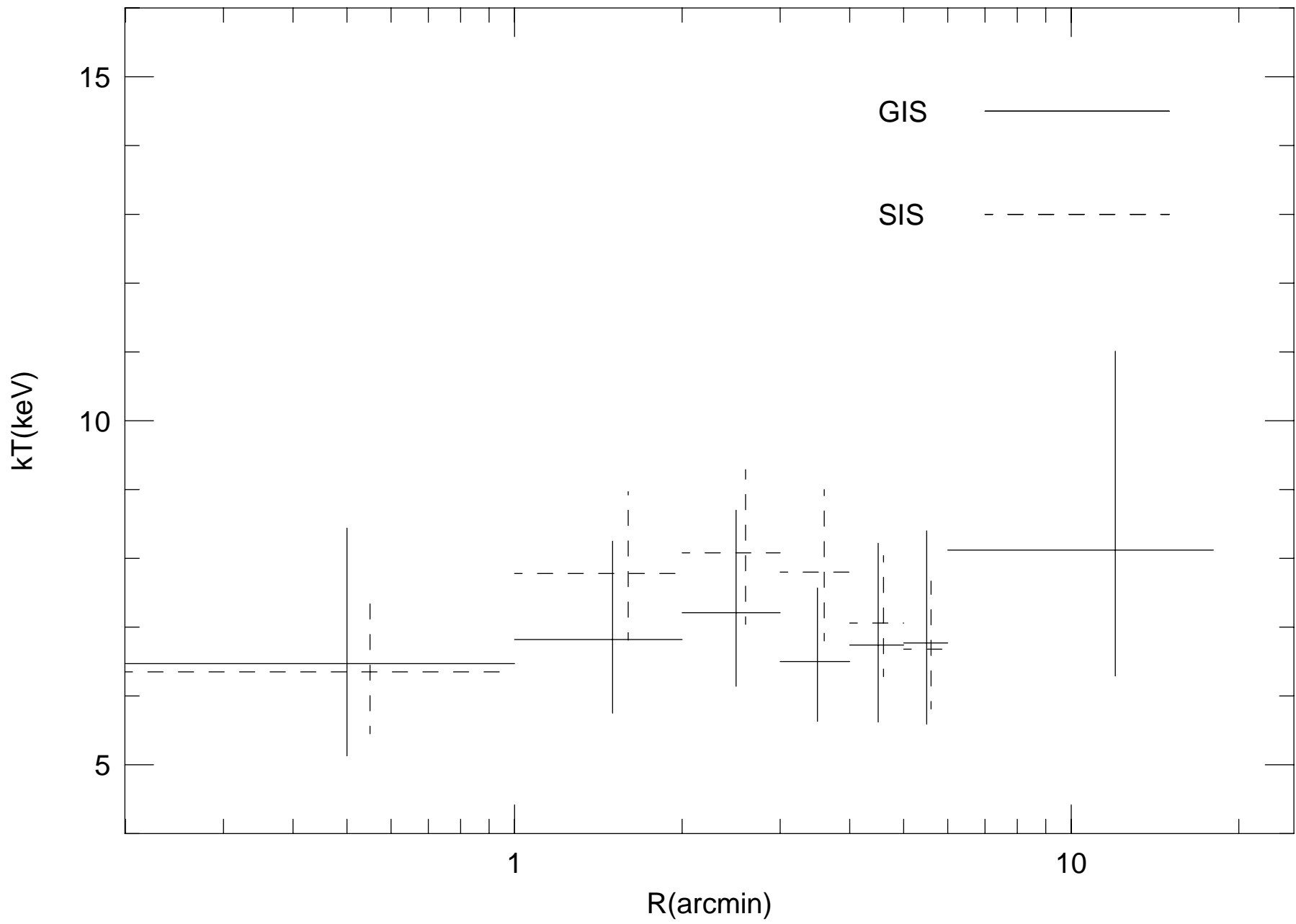
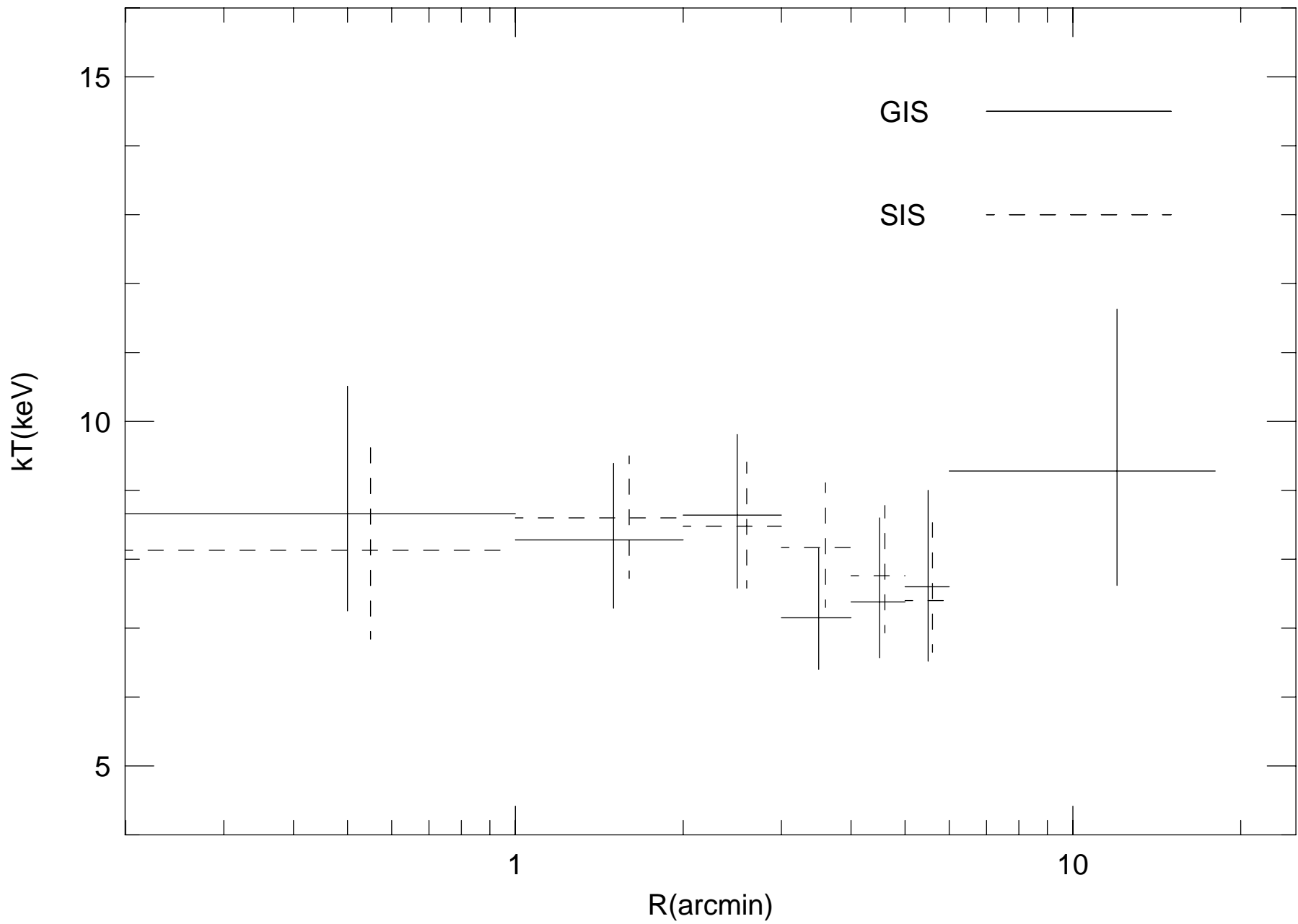
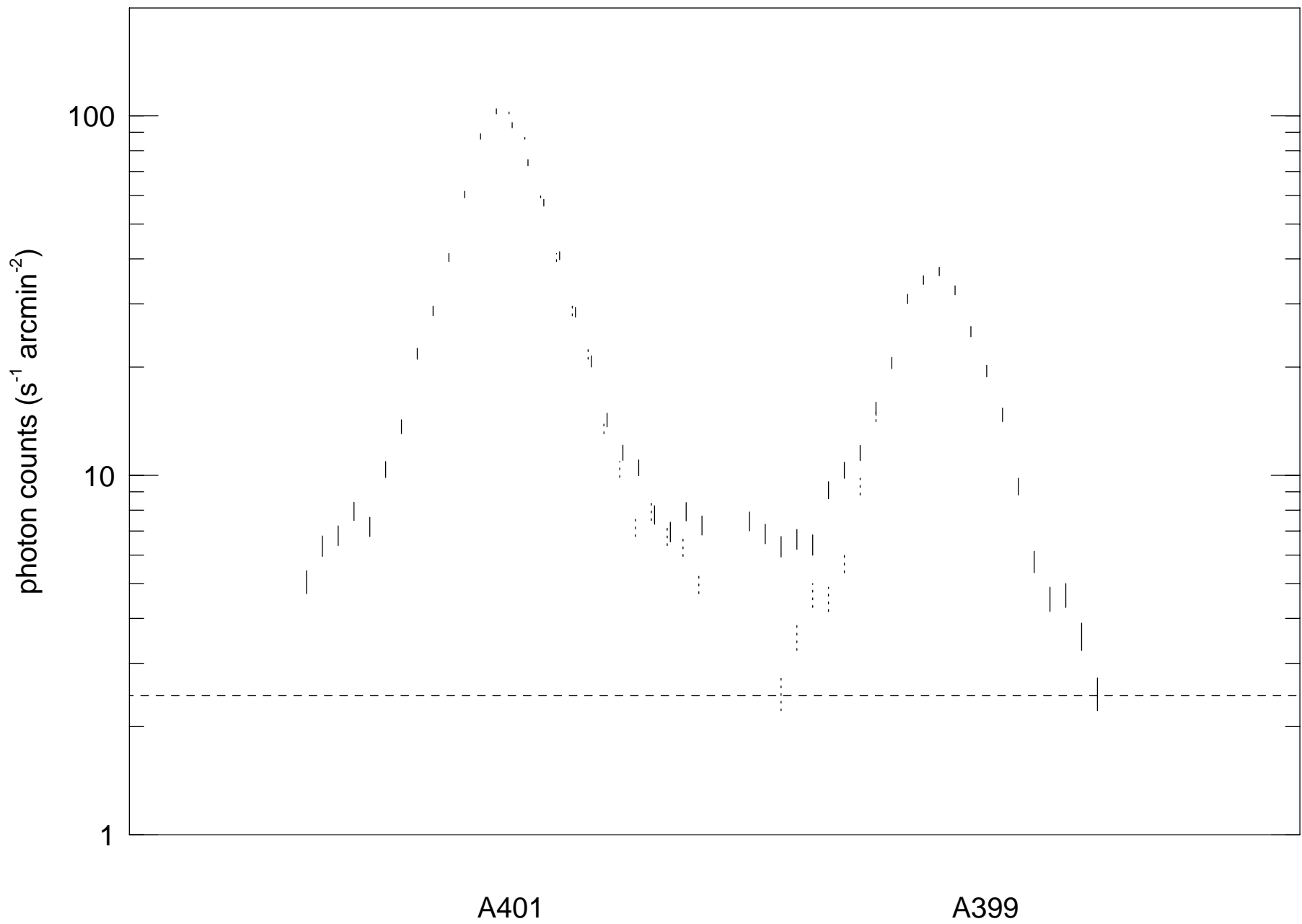
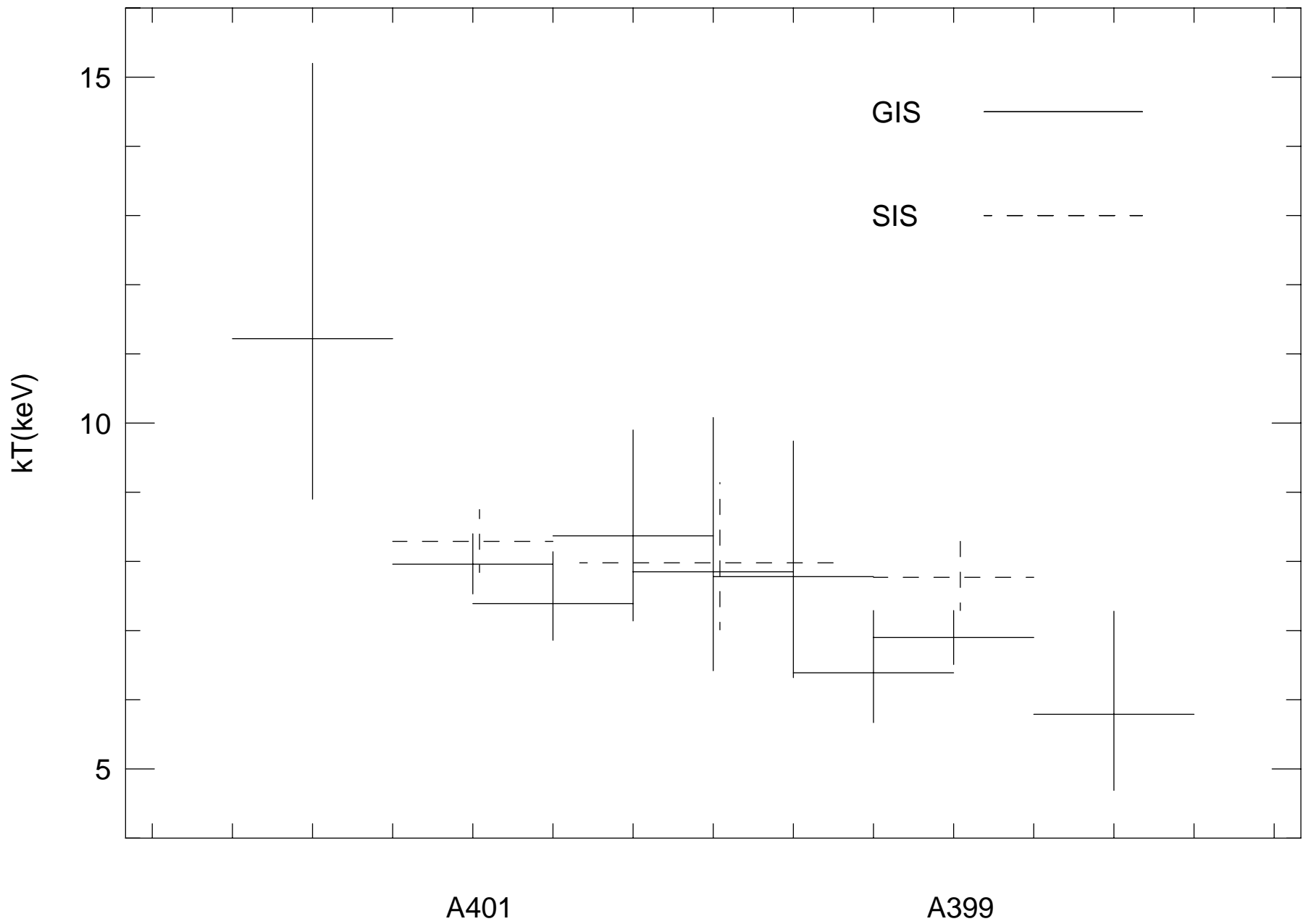
\end{document}